\documentstyle[aps,prc]{revtex}
\tightenlines
\begin{document}

\title{``Colliding beam" enhancement mechanism of deuteron-deuteron fusion 
reactions in matter}

\author{B.L. Altshuler$^{1}$,
 V.V.Flambaum$^{2}$\thanks{email address: flambaum@newt.phys.unsw.edu.au},
 M.Yu. Kuchiev$^{2}$ and 
V.G. Zelevinsky$^{3}$}

\address{$^1$ Physics Department, Princeton University, Princeton, NJ 08544
 and NEC Research 
Institute, 4 Independence Way, Princeton NJ 08540, USA}

\address{$^2$ School of Physics, University of New South Wales,
Sydney 2052, Australia}

\address{$^3$Department of Physics and Astronomy and National Superconducting
Cyclotron Laboratory, \\
Michigan State University, East Lansing, MI 48824-1321, USA}          
         
\date{\today}
\maketitle


\begin{abstract}
 We suggest a ``ping-pong'' mechanism of enhancement
 for fusion reactions between a low energy external deuteron beam
and the deuterons in a condensed matter or molecular target.
 The mechanism is based on the possibility
of acceleration of a target deuteron by
 the Coulomb field of a projectile deuteron with its subsequent rebound 
from a heavy atom in matter and the following fusion of the two deuterons 
moving towards each other. This effectively converts 
the fixed target process  
into a colliding beam reaction. In a simple limiting case this
reduces the negative penetrability exponent by a factor of $\sqrt{2}$.
We also discuss a contribution  given by ``zero
 oscillations'' of a bound target deuteron.
The proposed mechanism is expected to be efficient in compounds with 
target deuterons localized in the vicinity of  heavy atoms.
\end{abstract}

\pacs{PACS numbers: 25.60.Pj , 61.90.+d  }


  It is well known that the fusion at low energies is exponentially suppressed
due to the Coulomb repulsion. Consider, for example, two deuterons with
a relative velocity $v$  in their center-of-mass reference frame.
 The Coulomb barrier penetration factor $P(v)$ is given by \cite{Landau}
\begin{equation}
\label{S} P(v) \simeq \frac{2 \pi e^2}{\hbar v} \exp
\left(-\frac{2 \pi e^2}{\hbar v}\right)
\end{equation}
There are a few publications (see , for example, \cite{1,2,Pd,Ichimaru}
and references therein)
 that claim  a substantial enhancement
of the DD fusion cross-section in solids. The present paper does not aim
to interpret particular experiments.  We just analyze some effects
in condensed matter or molecules containing deuterons which  result in
 certain enhancement.

Consider a target experiment where one of the deuterons is at rest before
 the collision whereas another one moves with a velocity $v_0$. 
The total energy of the projectile and target deuterons $E_0$ at infinity
 can be decomposed into the
energy of relative motion $E_r$ and the energy of the motion
of the center-of-mass $E_{c}$,
\begin{equation}
\label{E} E_0=\frac{mv_0^2}{2}=E_r+E_c= \frac{\mu v^2}{2}+\frac{M v_c^2}{2}
\end{equation} 
where $m$ is the deuteron mass, $\mu=m/2$ is the reduced mass,
$M= 2m$ is the total mass, $v,v_c$  are  the relative velocity and the
velocity of the center-of-mass.
For a pure two-body problem the relative velocity equals $v=v_0$ and the 
velocity of the center-of-mass is $v_c=(v_0+0)/2=v_0/2$. Since
 the barrier penetration factor $P(v)$ is determined by 
the relative velocity $v$, half of the energy, the energy of
 the center-of-mass, is wasted.

Imagine now that as a result of an elastic interaction with , for example,
heavy atoms the system transfers part of its momentum ${\bf q}$ without
changing  its energy.
This process  changes the center-of-mass
velocity:
${\bf v_c}={\bf v_0}/2 - {\bf q}/2m$. Provided that
 ${\bf v}_0 \cdot {\bf q} >0$ the center-of-mass
velocity is reduced. According to the energy constrain eq. (\ref{E}) 
the relative velocity increases:
\begin{equation}
\label{v} 
v_{\bf q}= \left[2 {\bf v}_0^2 - ({\bf v_0}- {\bf q }/m)^2\right]^{1/2}~,
\end{equation} 
leading to an exponential enhancement of the penetration factor $P(v)$,
see eq. (\ref{S}). This enhancement can be understood in the
following way: as a result of the elastic interaction the two deuterons
move towards each other as they do in colliding beams.  
The maximal enhancement is achieved for $q=mv_0$. Under this condition
 $v({\bf q})$ is equal to $\sqrt{2} v_0$
instead of $v_0$ for free motion, i.e. the negative penetrability exponent
in eq. (\ref{S}) is reduced by a factor $\sqrt{2}$. Even a moderate
reduction of the center-of-mass momentum can dramatically increase
the fusion cross section.  

  How can the center of mass momentum be  reduced?
 There are two obvious ways to do this:
\itemize
\item Due to the uncertainty principle the target deuteron is never at rest.
The wave function $\phi ({\bf q})$ of its original state contains 
nonzero momentum ${\bf q}$ 
components.  Assuming that the target deuteron is confined within 
a harmonic potential well and thus behaves as a harmonic 
oscillator  with a frequency $\omega$, we have \cite{Landau} 
\begin{equation}
\label{phi0} \phi(q)= (\pi \omega \hbar m)^{-1/4}\exp(-q^2/2
\omega \hbar m)
\end{equation} 
Here we consider a one dimensional oscillator since
only the projection of the target deuteron momentum on the projectile
 momentum is important. 
 
The barrier penetration factor for fusion can be estimated as
\begin{equation} 
\label{factor} P_{{\rm eff}}= \int |\phi(q)|^2 P(v_q) 
dq=P(v_0) \exp\left(\frac{\pi^2e^4\omega}{mv_0^4\hbar}\right)~,
\end{equation}              
where $P(v_q)$ is given in (\ref{S}) and 
the velocity $v_q$ can be obtained expanding
(\ref{v})  as $v_q \approx v_0 + q/m$.
Thus, the ``zero  oscillations'' of the target deuteron give an enhancement
 factor
\begin{equation}
\label{enh} 
S=\exp\left(\frac{\pi^2e^4\omega}{mv_0^4\hbar}\right)
\end{equation}
  Note that in an excited oscillator state  $<q^2> \propto (2 n+1)$,
where $n$ is the oscillator quantum number. Correspondingly, the enhancement
factor $S$ in excited states is much larger.

\item Another possibility to modify the center-of-mass momentum
arises from the fact that the interaction between the deuterons is a
long-range Coulomb interaction which forces the target deuteron to move
long before fusion. As a result the target deuteron obtains an
additional momentum which it can transfer to the environment. In the simple
oscillator model this environment is represented by the harmonic potential
well. In other words, the interaction with the projectile deuteron excites
the harmonic oscillator. This phenomenon can be understood classically:
the projectile deuteron plays ping-pong with the environment using
the target deuteron as a ping-pong ball.

   Let us present a classical  estimate of this effect
in a simple model. Assume that both deuterons
are influenced by a harmonic oscillator well and interact
between themselves by the Coulomb force.
To produce  fusion the collision of the two deuterons should be 
characterized by an extremely small impact parameter. Assuming that the 
harmonic well is isotropic we can neglect the transverse degrees of
freedom and reduce the three dimensional problem 
of the collision of the projectile deuteron with the oscillator to the 
one-dimensional problem with all particles moving along the $x$-axis. 
The Newton equations for the  two interacting deuterons  are
\begin{equation} \label{121}
m \frac{d^2x_t}{dt^2}=-\frac{dU(x_t)}{dx_t}+\frac{e^2}{|x_t-x_p|^2}
\end{equation}
\begin{equation} \label{122}
m \frac{d^2x_p}{dt^2}=-\frac{dU(x_p)}{dx_p}-\frac{e^2}{|x_t-x_p|^2}
\end{equation}
Here $U(x)$ is the external potential acting on the deuterons.
At large distances these equations describe  the motion of the projectile with 
a constant velocity $x_p=v_0 t$ and oscillatory motion of the target
near the bottom of the well. Substituting $|x_t - x_p| \approx
v_0t$ in the Coulomb potential energy we find
\begin{equation} 
\label{displacement}
x_t(t)=-\frac{e^2}{v_0^2 m} \int_{-\infty}^{\omega t}\frac{d\tau}{\tau}
\cos(\tau - \omega t  )
\end{equation} 
\begin{equation} 
\label{velocity}
v_t(t)=\frac{e^2}{v_0^2 m}\left(\frac{1}{|t|}-\omega
 \int_{-\infty}^{\omega t}\frac{d\tau}{\tau}
\sin( \tau -\omega t) \right)
\end{equation} 
The first term in eq.(\ref{velocity}) 
describes the motion in the absence of the oscillator
potential and corresponds to the momentum transfer between the deuterons.
 
We see that at large $|t|$ the velocity of the target deuteron $v_t$
oscillates; each of the oscillations corresponds to a certain momentum
transfer to the environment. The total momentum transfer to the environment
 can be estimated by taking the limit $t=0$ in the integral
 in eq.(\ref{velocity}):
\begin{equation} 
\label{transfer}
q=\frac{\pi e^2 \omega}{2v_0^2}
\end{equation} 
Substitution of this momentum transfer into eqs.(\ref{S},\ref{v})
gives the enhancement factor $S$ which is exactly equal to the
 contribution of the zero oscillations, eq.(\ref{enh}). A naive
addition of these two effects would give
 an enhancement factor
\begin{equation}
\label{enh1} 
S=\exp\left(\frac{2\pi^2e^4\omega}{mv_0^4\hbar}\right)
\end{equation}
( this result is equivalent to the increase of $<q^2>$ by a factor of two;
note that in the first excited oscillator state  $<q^2>$ is three
 times larger than that in the ground state.)
However, this estimate of the oscillator excitation contribution
 contains the approximations which can hardly
 be justified.  Indeed, the velocity of the projectile may be taken constant
at large distances only. Also, at large distances there is the screening 
of the deuteron interaction by electrons. Below we present the estimate
of the small distance contribution which seems to be more important.

It is convenient to rewrite the deuteron motion equations  in terms of
 the motion of the center-of-mass, $x_c=(x_t+x_p)/2$, and relative motion,
 $x_r=x_t-x_p$. Taking the sum and the difference of equations
(\ref{121}) and (\ref{122}) we obtain
\begin{equation} \label{123}
2m \frac{d^2x_c}{dt^2}=-\frac{dU(x_t)}{dx_t}-\frac{dU(x_p)}{dx_p}
\end{equation}
\begin{equation} \label{124}
\frac{m}{2} \frac{d^2x_r}{dt^2}=\frac{e^2}{x_r^2}
-\frac12\left(\frac{dU(x_t)}{dx_t}-\frac{dU(x_p)}{dx_p}\right)
\end{equation}
If the target deuteron is close to a heavy atom the repulsive force
$-\frac{dU(x_t)}{dx_t}$ acting on the target is larger than
the similar force $-\frac{dU(x_p)}{dx_p}$ acting on the projectile.
This means that the net ``external'' force acting on the relative
motion  opposes the Coulomb repulsion and increases the energy
of the relative motion! In fact, we have obtained the mechanism
to transfer the energy from the motion of the center-of-mass to the
relative motion. Note that the harmonic oscillator case is the special one.
 In this case we have the separation of the variables for the center-of-mass
motion and relative motion since
$kx_t^2/2+kx_p^2/2=kx_c^2+kx_r^2/4$. The gain in the energy of the relative   
motion in this case is about half of  the deuteron binding energy $E_b$
(in fact, in this model the energy gain is equal to
 the maximal value of $kx_r^2/4$ at the boundary of the finite size
 oscillator ). This gives the relative
velocity $v=\sqrt{2(E_0+E_b)/m}$ and the enhancement
factor 
\begin{equation}
\label{enh2} 
S=\exp\left(\frac{2\pi e^2 E_b}{mv_0^3\hbar}\right)
\end{equation}
 The energy gain can be larger if the deuterons collide
in the area of high repulsive potential produced by a heavy atom
 where the force $-\frac{dU(x_t)}{dx_t}+\frac{dU(x_p)}{dx_p}
 \sim 2 x_r  Z_{eff} e^2/r_c^3$ is large (here $r_c$ is the distance to the
 heavy nucleus, $Z_{eff}$ is the nuclear charge partly screened by the atomic
 electrons). Note, however, that at high energy $E_0$ of the projectile
 deuteron the displacement of the center-of-mass during the collision is
 relatively small:
 $\Delta x_c \sim  \frac{r_{min}}{4}(\ln(\frac{4 r}{r_{min}}) -1)$
where $r_{min}=2 e^2/E_0=2 a_B($27 eV/$E_0)$ is the classical  minimal distance
 between the deuterons and $r$ is the initial distance between them (it can
 be identified with the screening radius for the Coulomb interaction between
 the deuterons), $a_B$ is the Bohr radius.
For high energies, $\Delta x_c$ is small.  The deuterons just do not
 have enough time to reach the area of the high potential $U$
 before the fusion. In this case the collision process takes place close to the
equilibrium position of the target deuteron and we can approximate 
the potential $U$ by the oscillator one. Therefore, the estimate
 eq. (\ref{enh2}) seems to be
 a reasonable approximation for the  oscillator excitation effect
if the projectile energy is much larger than the atomic unit 27 eV . 

To present 
   numerical estimate for the zero oscillation  and excitation
 contributions let us assume that
$\hbar \omega = 0.3$ eV and $E_b= 5$ eV. Then
it follows from (\ref{enh}),(\ref{enh2}) that the enhancement factor
is
\begin{equation}\label{est}
S \sim \exp\left[  \left( \frac{300 eV}{E_0}\right)^2+
\left(\frac{200 eV}{E_0}\right)^{3/2}\right]~.
\end{equation} 
This estimate  shows that the mechanisms of the exponential
enhancement of the fusion  described above are efficient only
for the low energy deuterons with $E_0 < 1$ KeV . The low-energy enhancement
is larger for
 hard collisions (with momentum transfer $q \sim mv_0$) between the
 deuterons with rebound from heavy atoms.  The hard collisions
 can transfer all the center-of-mass energy to the relative motion.
However,
calculation of the hard collision  effects is quite cumbersome and
it will be published separately \cite{Kucheiv}.
 
In this paper we present the estimates for a deuteron beam experiment.
They can find further applications in 
non-equilibrium processes like chemical reactions, 
where cracks in solids or cavities in liquids with an electric field inside
can appear.
This electric field may accelerate deuterons and create the beam-like
situation. Some additional enhancement of the considered effect
may be due to the fact that the
target deuterons may  be in an excited state
in a non-equilibrium case. Another possibility may appear in ferroelectric
materials with  internal electric fields.

The beam-like problem may also
appear in a laser-induced fusion where ions are accelerated by the laser
field and the interaction with  electrons (see e.g. \cite{Hora}) .
We propose to add heavy atoms to the deuterium-tritium mixture.
The deuteron or tritium rebound from heavy
atoms produces a colliding beam and possible enhancement of the fusion.

Note that the density of matter during a laser-induced fusion  may be $10^4$
times higher than in usual solids (see, e.g. \cite{Hora1}), therefore, the
effects of environment are also much larger. For example, the fusion reaction
can happen in the area of
the strong potential of a heavy atom. As it was discussed above this leads
to the energy transfer from the motion of the center-of-mass to the relative
motion and exponential enhancement of the fusion probability. The effects of
the hard collisions are also very strongly enhanced (up to $10^8$ times).
 
Another important effect of  environment - partial screening of the
Coulomb barrier by  electrons - is discussed , for example, in Ref.
 \cite{Ichimaru}.

The authors are grateful to A. Galonsky who kindly introduced the
problem, and to  S. Billinge, P. Chandra, H. Hora, 
M. Kanatzidis, J. Scott, E. Shuryak, 
V. Sokolov, D. Tomanek, and M. Treacy for useful discussions.       
The work was supported by the Australian Research Council.


\begin{thebibliography}{99}
\bibitem{Landau}L.D. Landau and E.M. Lifshitz, {\sl Quantum mechanics:
non-relativistic theory}, Pergamon, New York, 1965.
\bibitem{1} S.E. Jones, E.P. Palmer, J.B. Czirr, D.L. Decker, G.L. Jensen, 
J.M. Thorne, S.F. Taylor, and J. Rafelski, Nature {\bf 338}, 737 (1989).
\bibitem{2} A. Arzhannikov and G. Kezerashvili, Phys. Lett. A{\bf 156}, 514
(1991).
\bibitem{Pd}H. Yuki, J. Kasagi, A.G. Lipson, T. Ohtsuki, T. Baba, T. Noda, 
B.F. Lyakhov, and N. Asami. Pis'ma Zh. Eksp. Teor. Fiz. {\bf 68}, 785 (1998)
[JETP Lett. {\bf 68}, 823 (1998)].
\bibitem{Ichimaru} S. Ichimaru. Rev. Mod. Phys. {\bf 65}, 255 (1994). 
\bibitem{Kucheiv} M.Yu. Kucheiv, B.L. Altshuler, V.V. Flambaum. Nucl-th/0010027.
\bibitem{Hora} H. Hora. Physics of Laser Driven Plasmas
 (John Wiley\&Sons, New York, 1981).
\bibitem{Hora1} H. Hora and R.J. Stening. The Physicist {\bf 36}, 220 (1999).
\end{thebibliography}
\end{document}